\documentclass[twocolumn,letter]{jpsj3}

\usepackage{amsmath,amssymb,bm}

\title{Low temperature properties of the fermionic mixtures 
with mass imbalance in optical lattice}

\author{Nayuta \surname{Takemori}
\thanks{E-mail address: takemori@stat.phys.titech.ac.jp} and 
\name{Akihisa \surname{Koga}}
}
\inst{\address{Department of Physics, Tokyo Institute of Technology, 
Megro, Tokyo 152-8551, Japan}
}
\abst{
We study the attractive Hubbard model with mass imbalance to clarify 
low temperature properties of the fermionic mixtures in the optical lattice.
By combining dynamical mean-field theory 
with the continuous-time quantum Monte Carlo simulation,
we discuss the competition between the superfluid and density wave states
at half filling.
By calculating the energy and the order parameter for each state,
we clarify that the coexisting (supersolid) state, where the density wave and 
superfluid states are degenerate,
is realized in the system.
We then determine the phase diagram at finite temperatures.
}

\kword{superfluid, density wave, mass imbalanced system, 
continuous-time Monte Carlo simulation}

\begin{document}
\maketitle
Superfluid state in ultracold fermionic systems has
attracted considerable interest
since the successful realization of 
the Bose-Einstein condensation in $\rm ^6Li_2$ molecules.\cite{BEC1,BEC2}
Due to the high controllability in the system, 
remarkable phenomena have been observed such as 
the BCS-BEC crossover\cite{BCSBEC1,BCSBEC2}
and the superfluid state in the spin-imbalanced system,\cite{Imbalance1,Imbalance2}
where Cooper pairs are composed of ions with distinct hyperfine states.
Recently, the fermionic mixtures with distinct ions, 
{\it eg.} $^6$Li and $^{40}$K,
have experimentally been realized,\cite{Taglieber08,Wille08}
which stimulates further theoretical
investigations on the superfluid states in the mass imbalanced system.
\cite{Liu,Iskin,Wu,Lin,Orso,Conduit,Guo,Baarsma,Diener}

One of interesting questions in such a mass imbalanced system
is how the superfluid state is realized 
when the lattice potential is loaded,
so-called, an optical lattice.\cite{Bloch}
In the lattice system,\cite{Dao} the density wave (DW) state is naively expected, 
in addition to the SF state,
since less mobile fermions tend to crystallize in the lattice,
particularly, at half filling.
It is desired to systematically discuss how the SF state
competes or coexists with the DW state in the optical lattice system. 
This topic is closely related to an important issue 
in condensed matter physics, so-called,
the supersolid state,\cite{He4,Wessel,SuzukiKawashima,Yamamoto,Super} 
since the DW state can be regarded as a sort of the solid state.
Therefore, the optical lattice system with the mass imbalance should be
providing an ideal stage for the studies of the supersolid state 
in fermionic systems.

Motivated by this, 
we study low temperature properties in the fermionic mixture 
in the optical lattice,
combining dynamical mean-field theory(DMFT)~\cite{Georges96} 
with the continuous-time quantum Monte Carlo (CTQMC) 
simulations~\cite{Werner,Werner11}.
By calculating the order parameters for the DW and SF states,
we determine the phase diagram at finite temperatures
and clarify how the coexisting state is stabilized against the mass imbalance.

In this paper, we consider the following attractive Hubbard model
with different masses,\cite{Dao} as
\begin{equation}
\label{eq:1}
H=\sum_{\langle i,j \rangle\sigma} t_{\sigma} (c^{\dagger}_{i \sigma} c_{j \sigma}+ {\rm h.c.})-U\sum_in_{i \uparrow}n_{i \downarrow}
\end{equation}
where $\langle i,j\rangle$ denotes nearest neighbor site, 
$c_{i\sigma}^\dag (c_{i\sigma})$ is the creation (annihilation) operator 
of a fermion at the $i$th site with spin $\sigma (=\uparrow, \downarrow)$ and 
$n_{i\sigma}=c_{i\sigma}^\dag c_{i\sigma}$.
$U(>0)$ is the attractive interaction and
$t_\sigma$ is the hopping amplitude for the fermion with spin $\sigma$,
where the effect of the mass imbalance is taken into account.

We examine low temperature properties of this model 
by means of DMFT\cite{Georges96} which maps the lattice
model to the problem of a single-impurity connected
dynamically to a "heat bath". 
The Green's function is obtained via the self-consistency condition imposed
on this impurity problem. 
When the DW and SF instabilities are equally treated in the DMFT framework,
the self-consistency equation for the sublattice $\alpha [=A, B]$
is given as\cite{Dao}
\begin{eqnarray}
\left[{\hat G}_{0\alpha}(z)\right]^{-1}&=&z {\hat\sigma}_0+
\mu {\hat\sigma}_z - \frac{1}{4}{\hat T}{\hat G}_{\bar{\alpha}}(z){\hat T},
\end{eqnarray}
where ${\hat\sigma}_0$ is the identity matrix, 
${\hat\sigma}_z$ is the $z$-component of the Pauli matrix.
$\mu$ is the chemical potential and
${\hat T}={\rm diag}\left(D_\uparrow, -D_\downarrow\right)$,
where $D_\sigma$ is the half bandwidth for the bare band with spin $\sigma$.
${\hat G}_{0\alpha}(z)$ and ${\hat G}_\alpha(z)$ 
are the noninteracting Green function for 
the effective impurity model and the local Green function for 
the sublattice $\alpha$, which are represented in the Nambu formalism.

There are various methods
to solve the effective impurity problem. 
To study how the competition between the DW and SF states 
in the mass imbalanced system, 
an unbiased and accurate numerical solver is
necessary, such as the exact diagonalization or the
numerical renormalization group. 
A particularly powerful method for exploring finite temperature properties
is the hybridization-expansion CTQMC method,
\cite{Werner,Werner11}
which enables us to study the attractive Hubbard model both 
in the weak- and strong-coupling regimes.\cite{Koga11}
In the paper, by varying the ratio of the bandwidths $r=D_\downarrow/D_\uparrow$
with a fixed $D_\uparrow=1$ (energy unit), 
we proceed to discuss how the mass imbalance affects low temperature properties.

In the mass balanced system ($r=1$), 
low-energy properties have been studied 
in one dimension,\cite{Lieb,Shiba,Machida,Fujihara}
two dimensions\cite{TD1,TD2} and 
infinite dimensions.\cite{Freericks,Suzuki,Keller,Capone,Garg,Bauer,Koga10}
It is known that 
the DW and SF states are 
degenerate on the bipartite lattice in two and higher dimensions
at half filling.
Each order parameter can be defined by
\begin{eqnarray}
\Delta_{DW}&=&\frac{1}{N}\sum_{i\sigma}(-1)^i\langle n_{i\sigma}\rangle,\\
\Delta_{SF}&=&\frac{1}{N}\sum_i\langle c_{i\uparrow} c_{i\downarrow}\rangle.
\end{eqnarray}
In this case, the supersolid state with both order parameters is possible 
to be realized.
On the other hand, in the imbalanced limit ($r=0$),
the system is reduced to the spinless Falicov-Kimball model
with mobile and localized fermions,\cite{FK,BM,FreerickRMP,FKCDW}
where the ground state is the genuine DW state.
Therefore, it is necessary to clarify how the introduction of the 
mass imbalance lifts the degeneracy of these two states.

To clarify this, we first discuss the stability of each state in the system.
\begin{figure}[htb]
\begin{center}
\includegraphics[width=8cm]{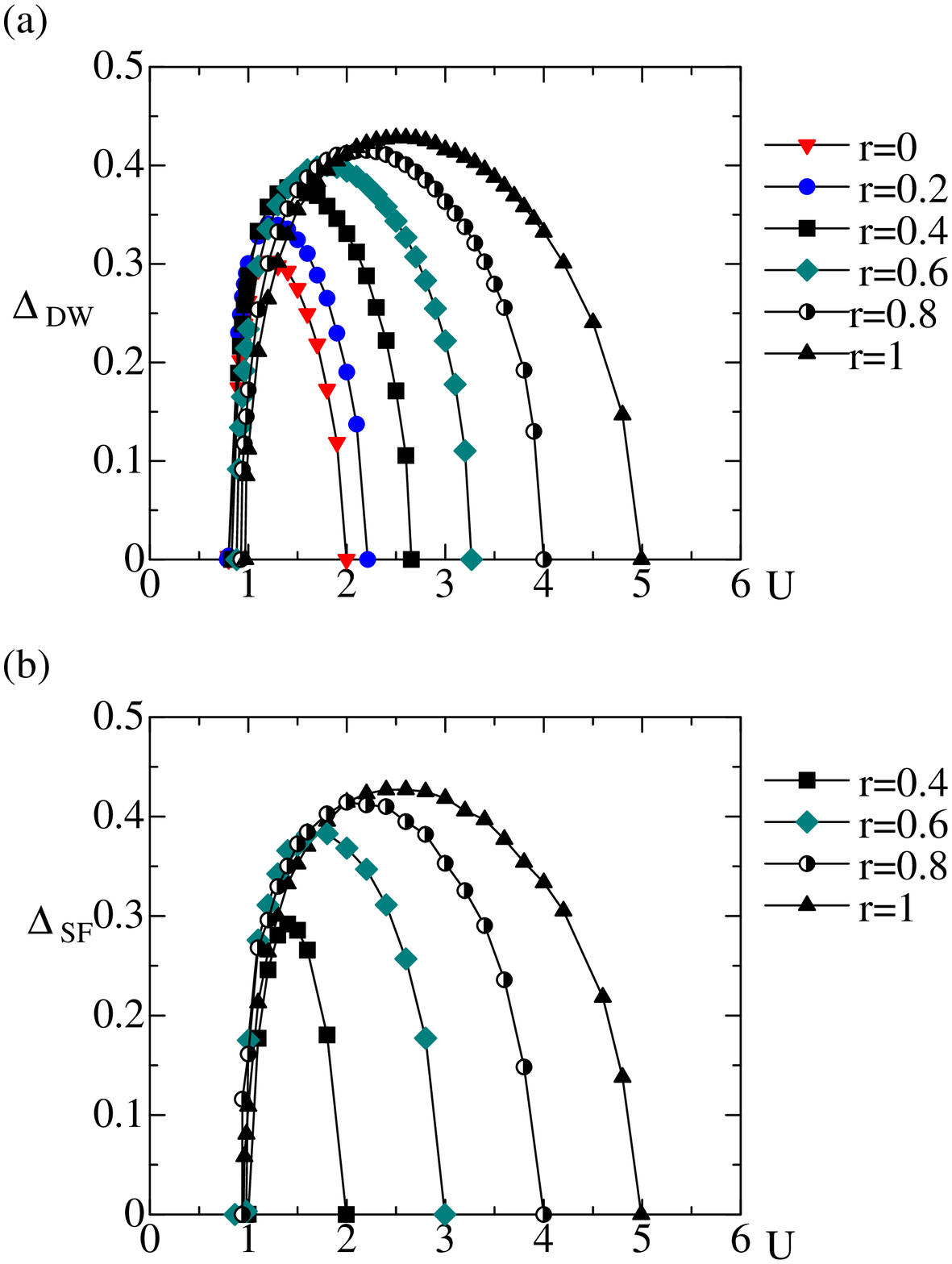}
\end{center}
\caption{
Order parameters for density wave (a) and superfluid (b) states 
as a function of the interaction strength at $T=0.05$.
}
\label{f1}
\end{figure}
We calculate the order parameters $\Delta_{DW}$ and $\Delta_{SF}$
under the conditions $\Delta_{SF}=0$ and $\Delta_{DW}=0$, respectively.
The obtained results for the DW and SF states at $T=0.05$ 
are shown in Fig. \ref{f1}. 
In the mass balanced case ($r=1$), 
the system has the $SU(2)$ symmetry and thereby
we find that these two order parameters are identified 
within the numerical accuracy.
In the case, the nature of the phase transition is well-known.\cite{Freericks}
In the noninteracting case $U=0$, a normal metallic state is realized.
Increasing the interaction beyond $U_{c1}(=0.95)$
the order parameters are induced, and the phase transition occurs 
to the DW or SF state.
Further increase in the interaction drives the system to 
the normal metallic state at $U_{c2}(=5.0)$.

The introduction of the mass imbalance leads to different behavior where
$\Delta_{DW}\geq \Delta_{SF}$,
as shown in Fig. \ref{f1}.
When $r=0.4$, the critical interactions for both states are deduced as 
$U_{c1}(DW)=0.83$, $U_{c2}(DW)=2.6$, 
$U_{c1}(SF)=0.96$, and $U_{c2}(SF)=2.0$,
by examining the critical behavior $\Delta \sim (|U-U_c|/U_c)^\beta$ 
with the exponent $\beta=1/2$.
These results mean that the DW state is stable in the larger parameter space while
the SF state becomes unstable.
In fact, when $r<0.3$, the SF solution no longer exists
and the genuine DW state is realized in the intermediate coupling region.

\begin{figure}[htb]
\begin{center}
\includegraphics[width=8cm]{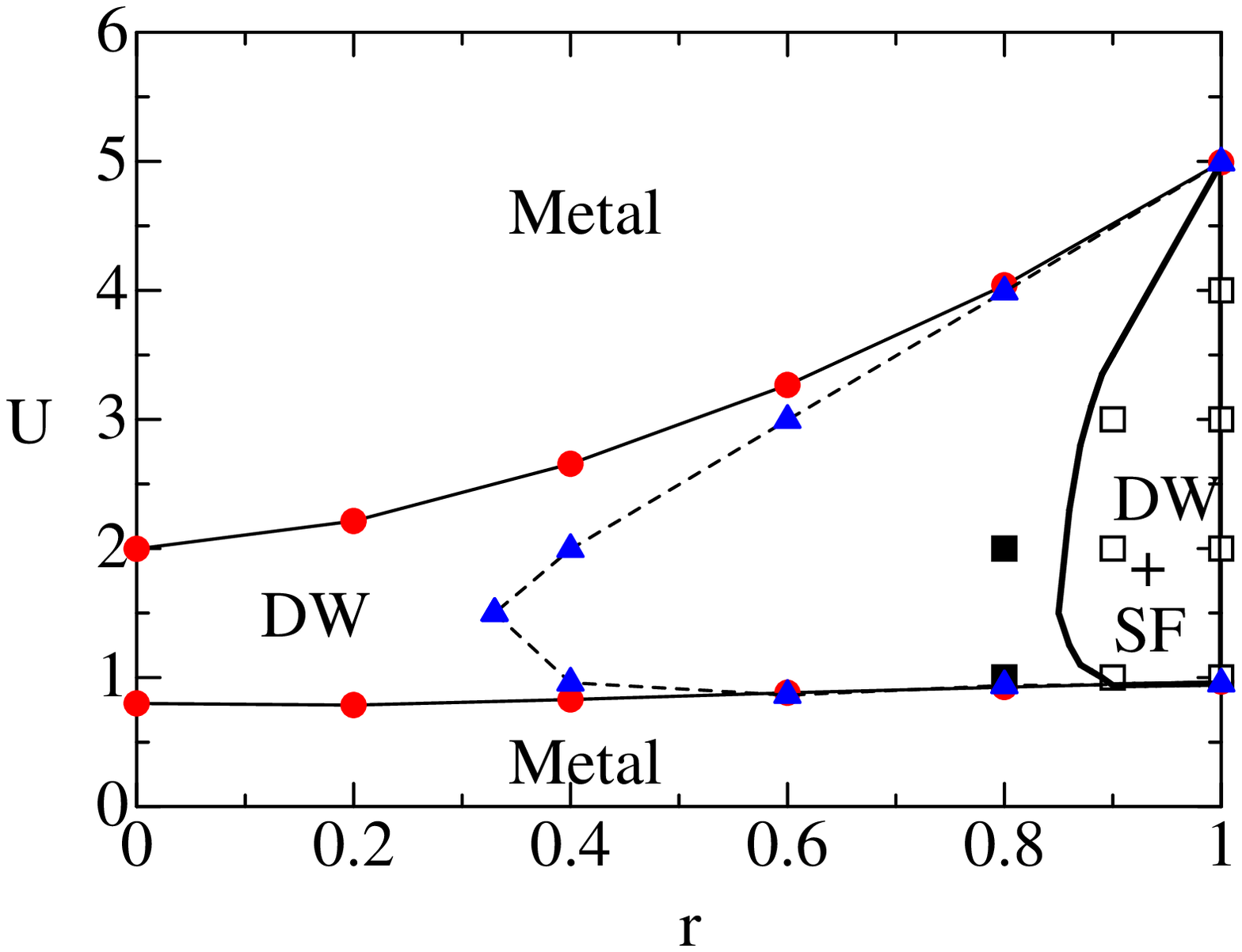}
\end{center}
\caption{
Phase diagram of the half-filled attractive Hubbard model at $T=0.05$,
where the lines are guides to eyes.
Solid lines represent the phase boundaries and
a dashed line represents the boundary where the unstable SF solution disappears.
Open (solid) squares indicate the coexisting (genuine DW) state.
}
\label{f2}
\end{figure}

\begin{figure}[htb]
\begin{center}
\includegraphics[width=8cm]{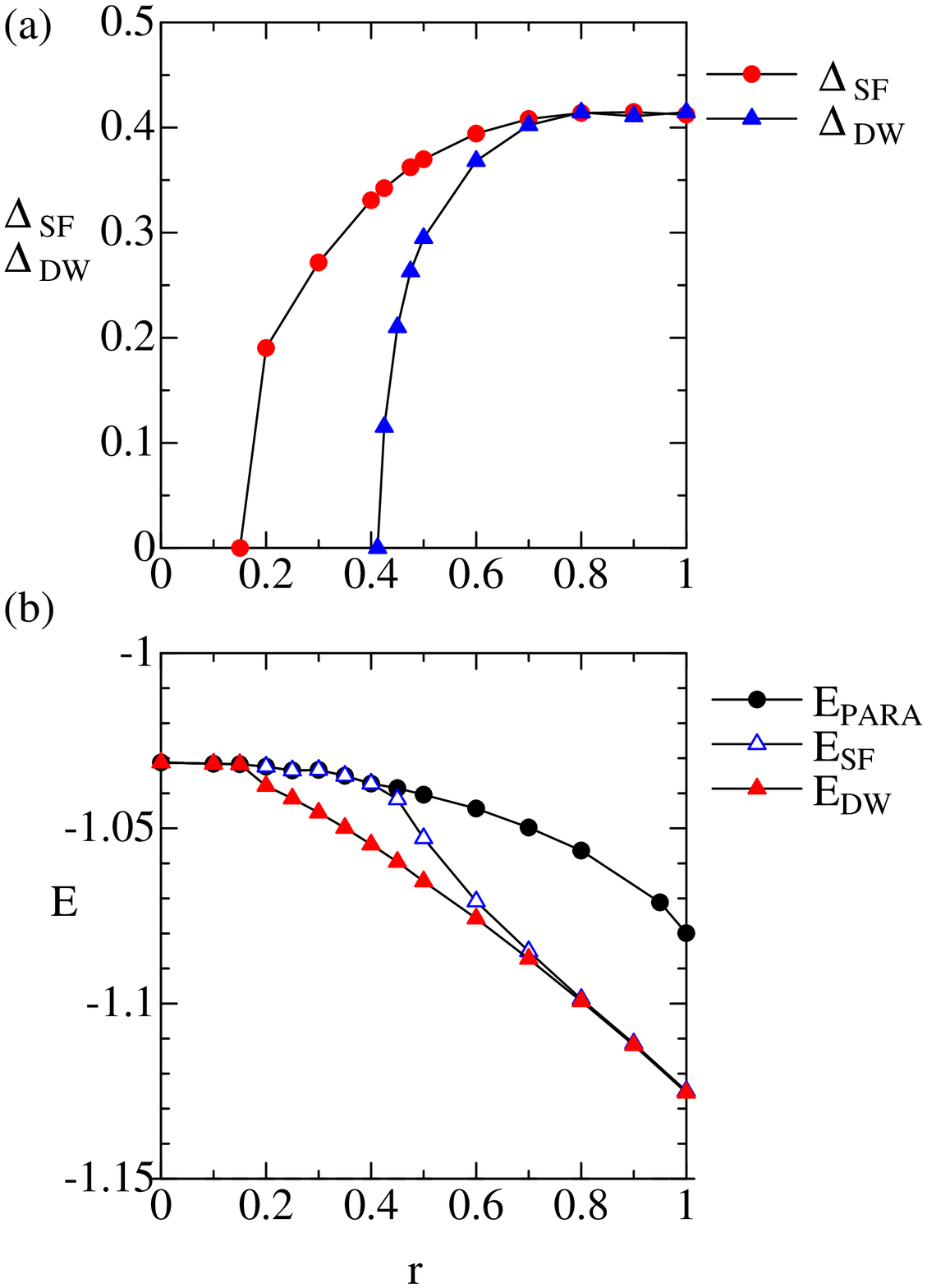}
\end{center}
\caption{Order parameters (a) and internal energies (b)
as a function of the ratio $r$ 
in the optical lattice with $U=2$ at $T=0.05$.
}
\label{f3}
\end{figure}

By performing similar calculations, we obtain the phase diagram at $T=0.05$, 
as shown in Fig. \ref{f2}.
In the weak and strong coupling regions, 
the normal metallic state is realized due to thermal fluctuations.
When $r=1$, the coexisting state appears in the region
$(U_{c1}<U<U_{c2})$, as discussed above.
Introducing the mass imbalance on the system with $U=2$,
the order parameters for the DW and SF states are monotonically decreased
 and at last disappear at the critical points $r_c^{DW}(\sim 0.15)$ and 
 $r_c^{SF}(\sim 0.42)$, as shown in Fig. \ref{f3} (a).
To study how the degeneracy of these two states is lifted, 
we also calculate the internal energy for each state
$E_{\rm DW}=E_K^{\rm DW}+E_U$ and $E_{\rm SF}=E_K^{\rm SF}+E_U$,\cite{Dao}
with
\begin{eqnarray}
E_K^{\rm DW}&=&\sum_{\sigma} \left(\frac{D_\sigma}{2}\right)^2\int_0^{\beta} d \tau 
G_{A\sigma}(\tau) G_{B\sigma}(- \tau),\\
E_K^{\rm SF}&=&\int _0 ^{\beta} d \tau 
\left[\sum_{\sigma} \left(\frac{D_\sigma}{2}\right)^2G_{\sigma}( \tau) G_{\sigma}( -\tau)\right.\nonumber\\
 &&- \left.\frac{D_\uparrow D_\downarrow}{4} F(\tau)F( - \tau)\right],\\
E_U&=&-\frac{U}{N}\sum_i \langle n_{i \uparrow}n_{i \downarrow}\rangle,
\end{eqnarray}
where $G(\tau)$ and $F(\tau)$ are the normal and anomalous Green's functions.
The results are shown in Fig. \ref{f3} (b).
It is found that when $r_c\leq r\leq 1$, 
the internal energies are identified 
within the numerical accuracy, where $r_c\approx 0.8\sim 0.9$.
Therefore, we can say that the coexisting state 
is realized in the region.
Below $r=r_c$, the degeneracy of two states is lifted, where
the genuine DW state is realized and the SF state becomes unstable.
Further decrease in the ratio $r$ yields the second-order phase transition 
to the normal state at the critical value $r_c^{DW}$, where
the order parameter vanishes and the curve of the energy  
for the DW state merges with the paramagnetic one.
In the limit $r=0$, the normal metallic state appears at $T=0.05$,
which is consistent with 
the result for the Falicov-Kimball model.\cite{FKCDW}

\begin{figure}[htb]
\begin{center}
\includegraphics[width=8cm]{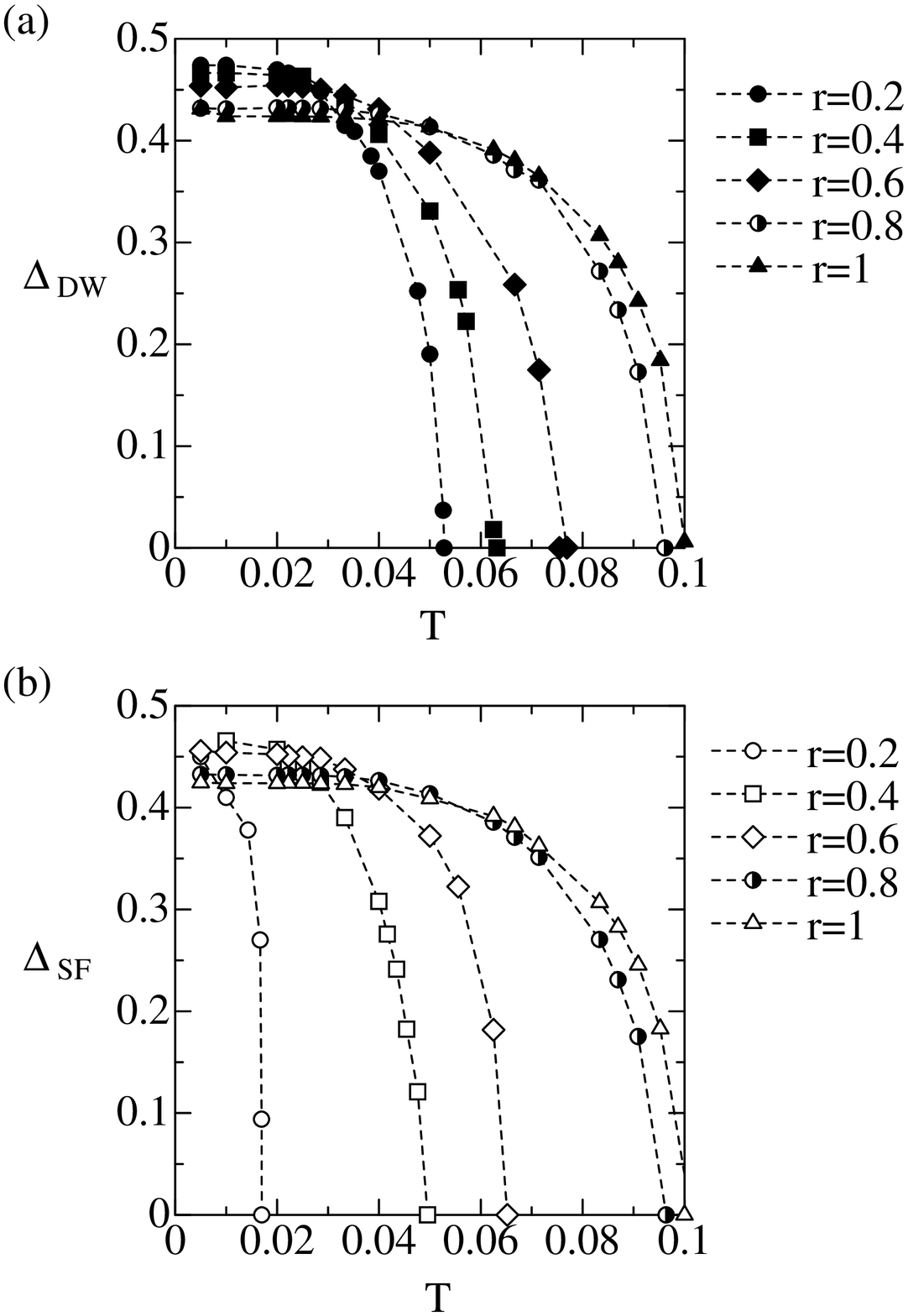}
\end{center}
\caption{Order parameters for the density wave (a) and superfluid (b) states
as a function of the temperature $T$ in the system with $r=0.2, 0.4, 0.6, 0.8$ 
and $1.0$, 
which are obtained under the conditions $\Delta_{SF}=0$ and $\Delta_{DW}=0$, 
respectively.}
\label{f4}
\end{figure}

Fig. \ref{f4} shows the temperature dependence of the order parameters
for the SF and DW states in the system with $U=2$,
which are obtained under the conditions $\Delta_{DW}=0$ and $\Delta_{SF}=0$, 
respectively.
We find that when $r_c\leq r \leq 1$, the magnitudes of these quantities 
are identified.
This implies that
the coexisting state is realized below the critical temperature.
On the other hand, in the small $r$ case,
the critical temperatures for the DW and SF states are different from each other,
{\it e.g.} $T_{c}^{SF}=0.065$ and $T_{c}^{DW}=0.076$ when $r=0.6$.
This implies that the SF state is unstable, and
the genuine DW state is realized below $T_c^{DW}$.


We wish to discuss ground state properties in the strong coupling limit.
To this end, 
we first use the particle-hole transformation\cite{Shiba} as
$a_{i\uparrow}=c_{i\uparrow}$ and $a_{i\downarrow}=(-1)^i c_{i\downarrow}^\dag$,
and map the negative-$U$ Hubbard model to the positive-$U$ Hubbard model.
Then, the effective Hamiltonian at half filling
is given by the XXZ quantum spin $s=1/2$ model as,
\begin{equation}
H = J \sum_{\langle i,j \rangle}\left[ {\bf S}_i \cdot {\bf S}_j 
+ \delta S_i^z \cdot S_j^z\right],
\end{equation}
where ${\bf  S}_i = \frac{1}{2}\sum a^\dag_{i\alpha} 
{\bf \sigma}_{\alpha\beta} a_{i\beta}$,
$\delta=(t_{\uparrow}- t_{\downarrow})^2/2 t_{\uparrow}t_{\downarrow}$ 
and $J=4t_{\uparrow}t_{\downarrow}/U$.
In the mass balanced case $(r=1)$, the effective model is reduced to the isotropic 
Heisenberg model.
The ground state is the antiferromagnetically ordered state, where
the direction of the ordered moment is arbitrary. 
Namely, when the ordered moments are along the $z$ axis (in the $x-y$ plane),
the DW (SF) state is realized in the original attractive Hubbard model.
On the other hand, the mass imbalance yields the anisotropy $\delta$ 
in the spin Hamiltonian, where the ordered moments are fixed along the $z$ axis.
This implies that in the strong coupling limit,
the coexisting state is realized only at $r=1$
and the genuine DW state is realized in general.
Therefore, we can say that in the attractive Hubbard model with different masses,
the phase boundary between the DW and coexisting states approaches the $r=1$ axis 
when $U$ increases.
The coexisting state is then realized only in the intermediate coupling region.
It is expected that the hole doping induces the genuine SF state 
(phase separation) in the coexisting (genuine DW) region of the phase diagram 
at half filling,
which is consistent with the results obtained from DMFT 
with the exact diagonalization.\cite{Dao}

Before closing the paper, we would like to comment on 
the realization of the supersolid state in the fermionic mixtures 
on the optical lattice.
We have confirmed that the coexisting state appears 
in the phase diagram.
It is expected that such interesting behavior appears in the three dimensional
optical lattice with the mass imbalance. 
However, one may consider that it is difficult to realize the supersolid state
 since SF fluctuations are hard to be controlled 
in the system.
Nevertheless, there are some possibilities to realize the supersolid state 
in the fermionic system.
One of them is the introduction of frustration 
to the DW state realized at low temperatures
since it tends to destabilize the DW state and to enhance 
SF fluctuations.
Another possibility is the introduction of the lattice potential 
to the superfluid state realized in the fermionic mixtures. 
By tuning DW and SF fluctuations by means of corresponding parameters properly,
the supersolid state should be realized 
in the fermionic mixtures on the optical lattice experimentally.

In summary, 
we have investigated low temperature properties of the fermionic mixtures
in the optical lattice,
which should be described by the attractive Hubbard model with different masses.
By combining DMFT with the strong-coupling version of the CTQMC method,
we have studied half-filling properties carefully to
clarify that the coexisting (supersolid) state, 
where the DW and SF states are degenerate, is realized.
By performing systematic calculations, we have obtained a rich phase diagram 
in the mass imbalanced attractive Hubbard model.

\begin{acknowledgment}
The authors thank P. Werner for valuable discussions. 
This work was partly supported by the Grant-in-Aid for Scientific Research 
20740194 (A.K.) and 
the Global COE Program ``Nanoscience and Quantum Physics" from 
the Ministry of Education, Culture, Sports, Science and Technology (MEXT) 
of Japan. The simulations have
been performed using some of the ALPS libraries.\cite{alps}
\end{acknowledgment}


\end{document}